\documentclass[conference]{IEEEtran}
\usepackage{blindtext, graphicx}

\usepackage[noadjust]{cite}
\usepackage{float}
\usepackage{amsmath}
\usepackage{graphicx}
\usepackage{amssymb}
\usepackage[colorlinks=true,citecolor=blue,pdftex]{hyperref}

\begin{document}

\title{RCD: Rapid Close to Deadline Scheduling for Datacenter Networks}

\author{
\IEEEauthorblockN{Mohammad Noormohammadpour$^{1}$, Cauligi S. Raghavendra$^{1}$, Sriram Rao$^{2}$, Asad M. Madni$^{3}$}
\IEEEauthorblockA{$^{1}$Ming Hsieh Department of Electrical Engineering, University of Southern California\\
$^{2}$Cloud and Information Services Lab, Microsoft\\
$^{3}$Department of Electrical Engineering, University of California, Los Angeles} 
\vspace{0.5em}
}

\maketitle

\begin{abstract}
Datacenter-based Cloud Computing services provide a flexible, scalable and yet economical infrastructure to host online services such as multimedia streaming, email and bulk storage. Many such services perform geo-replication to provide necessary quality of service and reliability to users resulting in frequent large inter-datacenter transfers. In order to meet tenant service level agreements (SLAs), these transfers have to be completed prior to a deadline. In addition, WAN resources are quite scarce and costly, meaning they should be fully utilized. Several recently proposed schemes, such as B4 \cite{b4}, TEMPUS \cite{calendaring}, and SWAN \cite{swan} have focused on improving the utilization of inter-datacenter transfers through centralized scheduling, however, they fail to provide a mechanism to guarantee that admitted requests meet their deadlines. Also, in a recent study, authors propose Amoeba \cite{amoeba}, a system that allows tenants to define deadlines and guarantees that the specified deadlines are met, however, to admit new traffic, the proposed system has to modify the allocation of already admitted transfers. In this paper, we propose Rapid Close to Deadline Scheduling (RCD), a close to deadline traffic allocation technique that is fast and efficient. Through simulations, we show that RCD is up to $15$ times faster than Amoeba, provides high link utilization along with deadline guarantees, and is able to make quick decisions on whether a new request can be fully satisfied before its deadline. 
\end{abstract}



\section{Introduction}

Cloud Computing \cite{cc} has changed the way IT services are provisioned, managed, and delivered to users. Companies do not have to over-provision costly hardware infrastructure and personnel to offer large scale services. Cloud Computing allows services to scale as needed and companies only pay for the amount of resources used. It can be considered as a pool of computing resources designed to provide ``a computing function as a utility" \cite{green}. In addition, through statistical multiplexing, public clouds, such as Amazon EC2 \cite{ec2}, are able to provide such flexible and scalable services at minimum costs. 

Datacenters are the infrastructure upon which cloud computing services are provided. To improve service provisioning, such infrastructures are usually spanned over multiple physical sites (multiple datacenters) closer to customers \cite{yahoo}. This also allows for higher reliability by keeping extra copies of user data on different datacenters. For example, Google's G-Scale network connects its 12 datacenters \cite{b4}. Also, Amazon EC2 platform runs on multiple datacenters located in different continents.  

Many applications require traffic exchange between datacenters to synchronize data, access backup copies or perform geo-replication; examples of which include content delivery networks (CDNs), cloud storage and search and indexing services. Most of these transfers have to be completed prior to a deadline which is usually within the range of an hour to a couple of days and can also be as large as petabytes \cite{calendaring}. In addition, links that connect these datacenters are expensive to build and maintain. As a result, for any algorithm used to manage these resources, it is desired to have the following features:

\textbf{Efficiency:} We want to maximize the transfer goodput. This means maximizing both link utilization and number of successfully finished transfers. 

\textbf{Speed:} For large scale applications that have millions of users, large number of transfers have to be processed and allocated. It is crucial that our algorithm is fast enough to allocate new requests in a short time. 

In \cite{d2tcp}, authors propose Deadline-aware Datacenter TCP (D2TCP) which increases the number of transfers that complete prior to their assigned deadlines by adjusting the transmission rate of such transfers based on their deadlines. Also, multiple previous studies have focused on improving the efficiency and performance of inter-datacenter communications through proper scheduling of transfers. In \cite{b4, swan, calendaring}, authors propose B4, TEMPUS and SWAN. B4 and SWAN focus on maximizing link utilization through centralized traffic engineering and application of software defined networking. TEMPUS improves fairness by maximizing the minimum portion of transfers delivered to destination before the transfer deadlines. None of these schemes guarantees that admitted transfers are completed prior to their deadlines. 

In \cite{amoeba}, authors propose Deadline-based Network Abstraction (DNA) which allows tenants to specify deadlines for transfers, and a system called Amoeba which guarantees that admitted transfers are completed prior to the specified deadlines. When a request is submitted, Amoeba performs admission control and decides whether the new request can be satisfied using available resources. If a transfer cannot be completed prior to its deadline, Amoeba tries to reschedule a subset of previously admitted requests to push traffic further away out of the new request's window. The admission process is performed on a first-come-first-served (FCFS) basis and requests are not preempted. In addition, each request is initially scheduled to finish as soon as possible. 

In this paper, we propose RCD: a simple and fast traffic scheduler that minimizes the time required to perform the admission process. RCD does not have to move already admitted requests to decide whether a new request can be admitted. Also, it achieves high utilization by effectively using resources. 

In the following sections, we first explain the problem of scheduling, present the rules based on which RCD operates, and compare RCD with Amoeba through simulations. 

\section{The Scheduling Problem}

As mentioned in several previous studies \cite{swan,calendaring}, inter-datacenter traffic can be categorized into three different classes: 

\textbf{Highpri:} This type of traffic is mainly initiated by users interacting with a website or online service. It is latency-sensitive as it directly affects user experience and has to be satisfied as soon as possible.

\textbf{Background:} Like previous traffic type, this one doesn't have a deadline either. It consists of throughput-oriented large transfers that are not critical, and have the least priority. 

\textbf{Elastic:} Traffic requests of this type have to be fully satisfied before their deadline. Deadlines can be either hard or soft. Finishing an elastic request with a hard deadline after its deadline is pointless while in a soft deadline case, value of the transfer drops based off a value function depending on how late the transfer is finished. 

Now imagine we have two datacenters $A$ and $B$ connected using one single link. New traffic requests of types mentioned above are generated in datacenter $A$ destined for datacenter $B$. Since highpri traffic is highly sensitive, it has to be sent on the link as soon as it is generated. The remaining bandwidth is first given to elastic requests and then background requests. The scheduling algorithm should decide how much bandwidth is given to each request at every time instance. 

It is possible to estimate the volume of highpri traffic (it is usually between 5-15\% of the link capacity) \cite{amoeba}. As a result, we can set aside a percentage of the bandwidth on the link for this traffic and solve the scheduling problem for elastic and background traffic considering only the left over capacity. Elastic traffic is always allocated first as it has a higher priority. Background traffic goes through when there is no more elastic traffic to send. Throughout this paper, we focus on Elastic traffic and ways of allocating it. 

Let us assume capacity $C$ is what is left after reserving the required bandwidth for highpri traffic. As in \cite{amoeba}, our aim here is to maximize the number of elastic transfers that finish before their deadlines as well as link utilization. In this paper, we will use linear programming to implement traffic scheduling.

In the following sections, we propose our method and compare its time complexity with the method proposed in \cite{amoeba}. Finally, we discuss how the ideas presented next can be further developed and applied to a network case where in addition to volume and deadline, each request is identified with a source and destination node. 

\section{Rapid Close to Deadline Scheduling}

In order to flexibly allocate traffic with varying rates over time, we break the timeline into small timeslots. We do not assume an exact length for these timeslots as there are trade-offs involved. Having smaller timeslots can lead to large linear programs (LPs) which are time-consuming and inefficient to solve while having larger timeslots results in a less flexible allocation because the transmission rate is considered constant over a timeslot. In implementation of the previous study \cite{amoeba}, a length of 3 to 5 minutes is used.

We schedule traffic for future timeslots because focusing only on current timeslot provides poor performance as we cannot provide long term promises and cannot guarantee that important transfers finish before their deadlines \cite{calendaring}. 
Assume we are allocating traffic for a timeline starting at $t_{now}$ representing current time and ending at $Td_{E}$ which corresponds to the latest deadline for all submitted requests. New requests may be submitted to the scheduler at any time. Each request is identified with three parameters, and is shown as $R(Q, Td_1, Td_2)$ in which $Q$ is the transfer volume, $Td_1$ is the first deadline where value of the transfer starts to drop if finished after this time, and $Td_2$ is the deadline after which finishing the transfer is pointless. For a request with a hard deadline $Td$, we have $Td = Td_1 = Td_2$. 

In order to schedule traffic, we can create and solve a linear program (LP) involving all of submitted requests with demand and capacity constraints populated based off link capacities and request volumes. This LP has to be solved every time a new request is submitted and can result in changing the allocation of already scheduled requests. The problem with this approach is its high complexity (solving such a large LP over and over is computationally inefficient) as the frequency of arrivals increases. 

In this section we discuss our proposal for bandwidth allocation. The useful property of elastic traffic is that it does not matter how soon or late we finish the transfer as long as we finish it before its deadline. We use this characteristic to further improve the allocation process based on the following rules:

\textbf{Rule 1:} Similar to previous schemes \cite{amoeba}, RCD does not support preemption. Preempting a request that is partly transmitted is wasteful. Also, it may result in thrashing if requests are consecutively preempted in favor of future requests. 

\textbf{Rule 2:} To be fast, RCD does not change the allocation of already allocated traffic unless there is leftover bandwidth in current timeslot ($t_{now}$). In which case it fetches traffic from the earliest timeslot that is not empty and sends it. This is done until either we fully utilize the current timeslot or there is no more elastic traffic to send.

In addition, we make the following simplifying assumptions. Study of the algorithm without these assumptions is left for future work.

\begin{itemize}
\setlength{\itemsep}{0.5em}
\setlength{\labelsep}{1em}

\item Requests are instantly allocated upon arrival. They are allocated over timeslots for which $t > t_{now}$. If a request cannot be allocated upon arrival, it is rejected. 

\item All requests have a hard deadline.

\item Bandwidth used by highpri traffic is constant and can be estimated with high accuracy.

\end{itemize}

\begin{figure*}[t]
\centering
\includegraphics[width=0.9\textwidth]{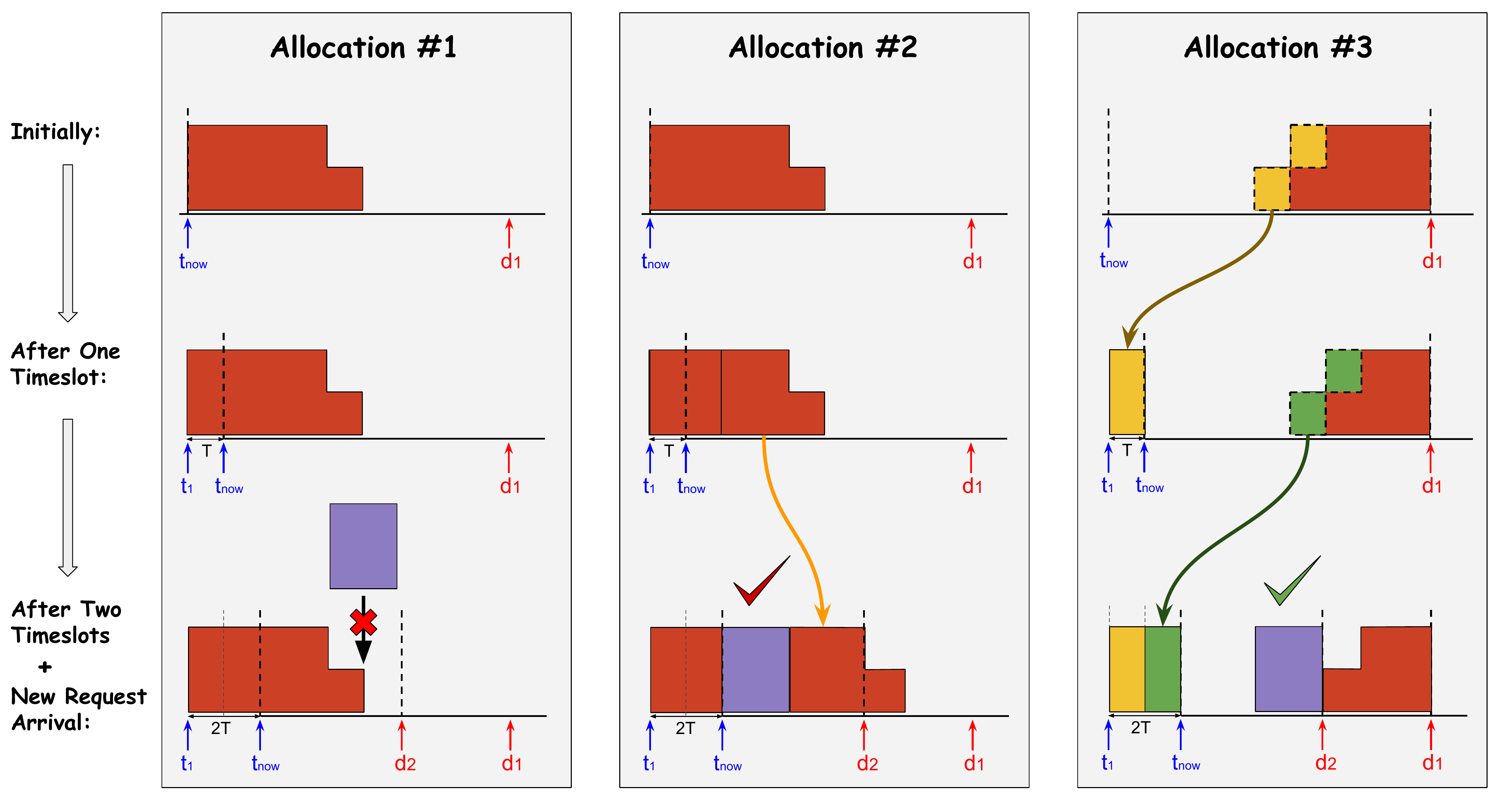}
\caption{Elastic traffic and different allocations}
\label{fig:timeline}
\end{figure*}

Now let's discuss an example considering our allocation rules. Fig. \ref{fig:timeline} shows three traffic allocation approaches and their behavior when a new traffic request is submitted:

\textbf{Allocation 1:} Tries to finish everything as soon as possible and does not change anything that is already allocated. This is the least complex approach; however, it fails to accommodate the new elastic request.  

\textbf{Allocation 2:} Tries to finish everything as soon as possible but upon arrival of the new request, moves part of the already allocated request ahead to allocate the new request. Depending on how many already allocated requests are moved to accommodate the new request, this method has a variable chance of success. If it reallocates a large number of requests then it can get very slow as the frequency of arrivals increases. Conversely, it can have a small chance of success if only a few requests are reallocated. 

\textbf{Allocation 3:} Based on our three rules, plans on finishing everything as late as possible, however, by sending traffic from future timeslots on $t_{now}$, it results in everything finished as soon as possible. Does not move allocated requests, only sends parts of already allocated requests to utilize the available capacity. Every time a new request is submitted, a small linear program involving only the new request is solved. 

When a new elastic request $E$ is submitted, RCD creates a small LP to schedule it. The size of this LP is $(Td - t_{now})$. Assume the amount of bandwidth allocated for which at time $t$ is $E(t)$, the total demand is $Q$, and the deadline is $Td$. In addition, let's assume $C_{t}$ is the residual capacity on the link at timeslot $t$. We use the LP of equation \ref{eq:eq2} with the objective function of equation \ref{eq:eq1} to do the allocation. If following LP does not yield a feasible solution, we reject the request.

Now assume elastic requests $R_1, R_2, ~... ~ R_K$ are submitted to the scheduler in order. We want to show that upon arrival of $R_k, ~ 1 \le k \le K$, our allocation for previously admitted requests is in a way that we cannot increase the admission chance of $R_k$ by rearranging the allocation of already allocated requests. Let's show the deadline of $R_k$ as $Td_k$ and at any time $t$, show the latest deadline of all admitted requests as $Td_E$. 

\begin{equation}
f(E) \triangleq \sum_{t=t_{now}+1}^{Td} E(t) \times (Td - t) \\
\label{eq:eq1}
\end{equation}

\begin{equation}
\begin{aligned}
 Obj: ~~ & min\{f(E)\} \\
 S.t: ~~ & \sum_{t = t_{now}+1}^{Td} E(t) = Q \\
 & 0 \le E(t) \le C_{t}, ~~~ t_{now} < t \le Td
\label{eq:eq2}
\end{aligned}
\end{equation}

\textbf{Theorem 1:} \textit{If we draw a vertical line at time $Td_k \le Td_E$ in our traffic allocation, it is not possible to increase the free space before the line by moving traffic from left side of the line ($t \le Td_k$) to the right side ($Td_E \ge t > Td_k$).}

\begin{figure*}[t]
\centering
\includegraphics[width=0.7\textwidth]{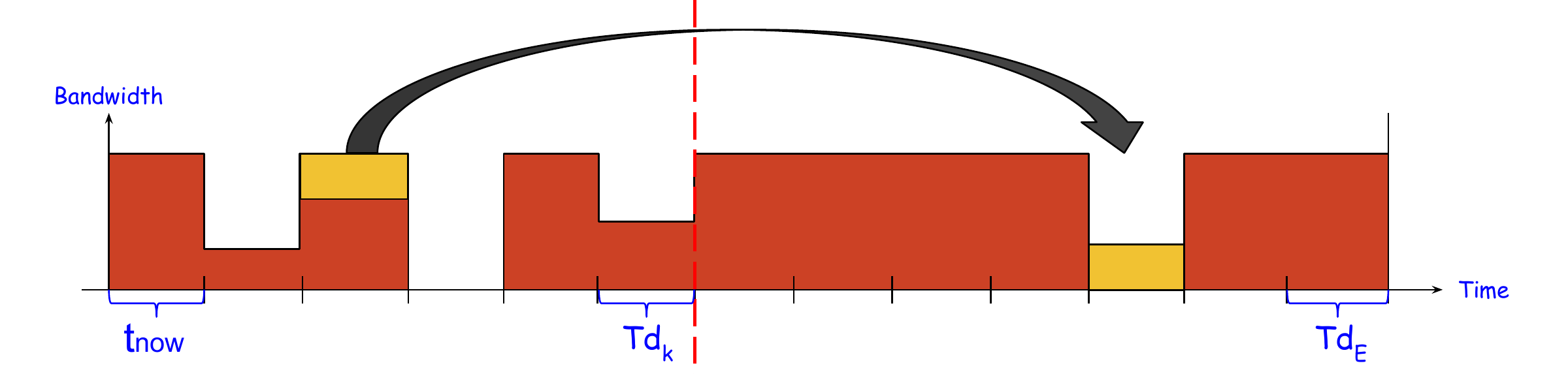}
\caption{A general allocation used in theorem 1}
\label{fig:theorem1}
\end{figure*}

\textbf{Proof:} Let's assume we have the allocation shown in Fig. \ref{fig:theorem1}. To schedule all requests, we used the cost function of equation \ref{eq:eq1} which assigns a smaller cost to future timeslots. Assume we can move some traffic volume from left side to the right side. If so, this volume belongs to at least one of the admitted requests and that means we are able to decrease the allocation cost for that request even further. This is not possible because the LP in equation \ref{eq:eq2} gives the minimum cost solution. Therefore, we cannot move traffic from left side of the line to the right side. Doing so will either result in violation of link capacity or violation of a deadline.  

Now let's assume a new elastic traffic arrives. If it can be allocated on the residual link capacity then we can accept it. If not, based on theorem 1, there is no way we can shift already allocated traffic so that we can accommodate the new elastic traffic. Therefore, we simply reject the request. 

\begin{figure}
\centering
\includegraphics[width=0.8\columnwidth]{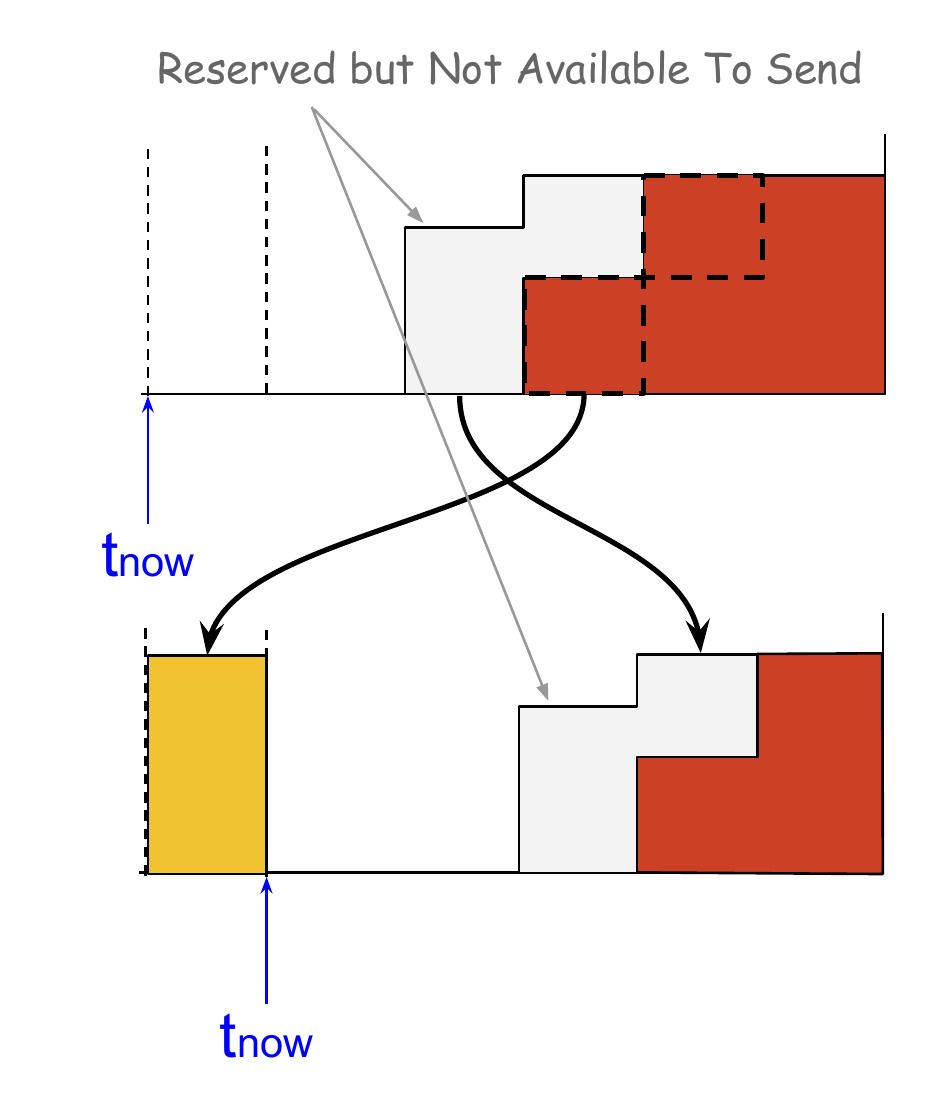}
\caption{Reserving bandwidth for a future request}
\label{fig:res}
\end{figure}

There may be cases when we want to reserve some bandwidth for an important traffic request while we don't have the traffic content yet. For example, at $6PM$ we find out an important traffic will arrive at $10PM$ and has to be fully transferred before midnight. We allocate such traffic the same as other elastic requests. This way we can be sure that when the traffic arrives, we have enough room to send it. If that important traffic is still not available by the time we finish sending all traffic requests that are allocated before it, we simply fetch traffic from the next closest request ahead of that transfer and if possible, push the important transfer to the next timeslot(s). Fig. \ref{fig:res} is an example of this case. 

If the reserved traffic is still not available when we are too close to its deadline that we cannot finish it anymore, we free the reserved bandwidth in favor of future elastic requests. 

\section{Simulation Results}

We compare the performance and speed of RCD with Amoeba \cite{amoeba}. Other schemes, such as \cite{swan, calendaring}, are deadline-agnostic and have an effective link utilization of less than $50\%$ \cite{amoeba}. Amoeba, on the other hand, only accepts requests when it can guarantee that the deadline can be fully met.

\textbf{Simulation Setup:} As mentioned earlier, we assume highpri traffic takes a fixed amount of bandwidth and allocate the leftover among elastic requests. Simulation is performed for 576 timeslots each lasting 5 minutes which is equal to 2 days. We performed the simulations three times and calculated the average.  

\textbf{Metrics:} Percentage of failed elastic requests, average link utilization, and average allocation time are the three metrics measured and presented. 

\textbf{Workload:} We generate elastic requests based off a Poisson distribution of rate $1 \le \lambda \le 8$. The difference between the arrival time of requests and their deadlines follows an exponential distribution with an average of $12$ timeslots. In addition, the demand of each request also follows an exponential distribution with an average of $0.286$ (a maximum of $1$ unit of traffic can be sent in each timeslot on the link). 

\begin{figure*}
\centering
\includegraphics[width=\textwidth]{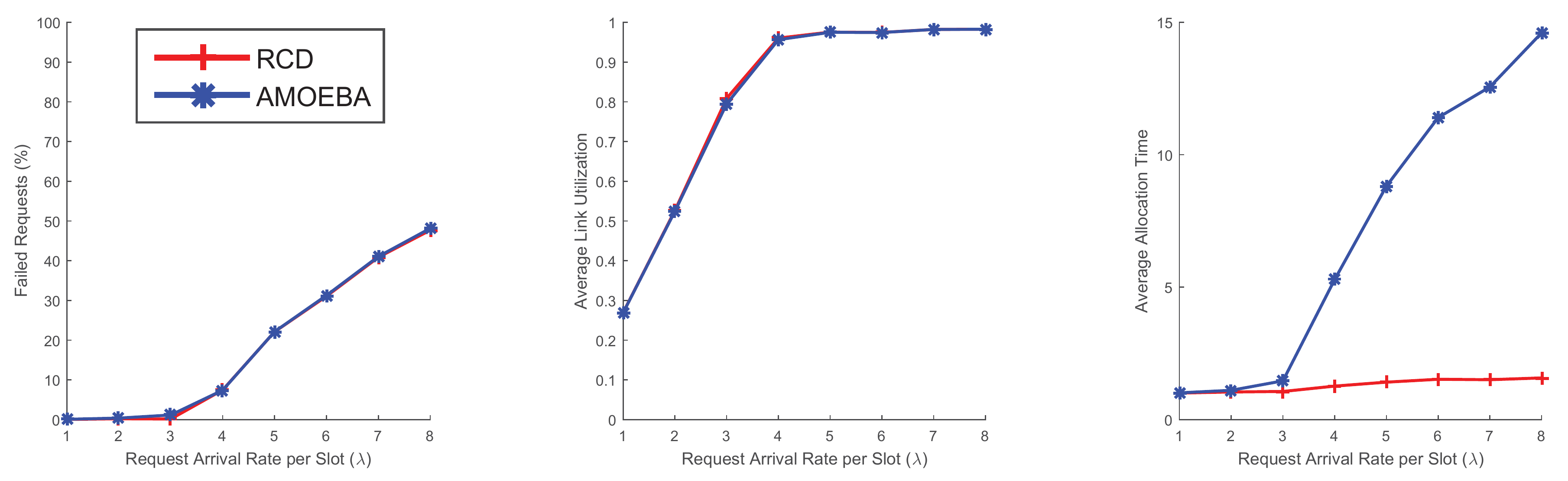}
\caption{Comparing Amoeba and RCD}
\label{fig:results_1}
\end{figure*}

Fig. \ref{fig:results_1} shows the aforementioned simulation metrics for both Amoeba and RCD. As can be seen, both algorithms result in similar failure rate and utilization. However, RCD achieves the same performance metrics with much less complexity: it is up to $15$ times faster than Amoeba. Also, the complexity of RCD grows very slowly as the frequency of arrivals increases: up to $1.6$ times while arrivals increase by a factor of $8$. 

With regards to the trend for time complexity as shown in Fig. \ref{fig:results_1}, when the request arrival rate is small, most of the capacity is left unused. Therefore, Amoeba does not have to move already allocated requests to push in a new one. As the arrival rate increases, we see a higher utilization. Starting the arrival rate of $4$, utilization gets close to $1$ and we can see a huge jump in the time complexity of Amoeba (by a factor of $3.7$). That is because Amoeba has to move around multiple already allocated requests to push in the new request. 

For an arrival rate of $8$ requests per timeslot, we see that both algorithms drop almost half of the requests. This can happen as a result of a failure in the network. For example, when a datacenter is connected using only two links and one of them fails for a few timeslots. While Amoeba can get really slow, RCD is able to handle such situations almost as fast as when there is low link utilization.

Up until now, the focus was only on one link. The network case will be discussed in the next section. 

\section{RCD in Network}

For a single link, it was demonstrated how RCD allows for independent allocation of newly arriving requests: one only needs to know the residual link capacity to allocate a new request, and can be sure that there will be no performance degradation. For a network case however, it is not as simple. In this section we discuss one possible way of extending RCD for application in a network connecting multiple datacenters. Providing a detailed solution is beyond the scope of this paper and is identified as future work. 

For a network case, each request is added a source and destination parameter and is shown as $R(Q, Td, S, D)$. We use the same rules as mentioned in section three. However, we have to augment the linear program in equation \ref{eq:eq2} to account for multiple paths and different capacity constraints of network links. Assume $L$ represents the set of network links $L_1, L_2, ~...~ L_m$ and $C_{t,l}$ is the residual capacity of link $l$ at time $t$. We want to allocate an elastic request with a source datacenter $A$ and a destination datacenter $B$. For a node $N$ in the network, $N_{out}$ refers to set of links going out of $N$ and $N_{in}$ refers to set of links going into $N$. We create the LP demonstrated in equation \ref{eq:eq4}, which corresponds to a flow network, with the objective function in equation \ref{eq:eq3} to allocate bandwidth upon arrival of a new request.

This LP has a size of $m(Td-t_{now})$ variables. For a network such as Google's G-Scale network \cite{b4} with 19 links, assuming 5-minute timeslots and a deadline of 24 hours, this LP will have 5500 variables. Using a laptop with Intel Core-i5 2.5GHz CPU, such an LP can be solved in less than a second. We can further optimize this LP using the k-shortest paths method: only links on those paths will be considered in routing. 

Based on the objective function in equation \ref{eq:eq3}, similar to single link case, the resulting allocation will favor later timeslots. However, this time traffic is initially scheduled as late as possible on whole paths (not each single link). It is straight-forward to show that theorem 1 still holds: assuming no preemption, if a new request arrives it is not possible to move any previously allocated traffic further into the future to open up space before new request's deadline.

\begin{equation}
f(E) \triangleq \sum_{t=t_{now}+1}^{Td} (\sum_{l=L_1}^{l=L_m} E(t,l)) \times (Td - t) \\
\label{eq:eq3}
\end{equation}

\begin{equation}
\begin{aligned}
 Obj: ~~ & min\{f(E)\} \\
 S.t: ~~ & \sum_{t=t_{now}+1}^{Td} \sum_{l \in A_{out}}^{} E(t,l) = Q \\
 & \sum_{t=t_{now}+1}^{Td} \sum_{l \in B_{in}}^{} E(t,l) = Q \\
 & \sum_{l \in N_{in}}^{} E(t,l) = \sum_{l \in N_{out}}^{} E(t,l), ~ t_{now} < t \le Td \\
 & 0 \le E(t,l) \le C_{t,l}, ~ t_{now} < t \le Td, \forall l~ \in~ L
\label{eq:eq4}
\end{aligned}
\end{equation}

\begin{figure}
\centering
\includegraphics[width=0.9\columnwidth]{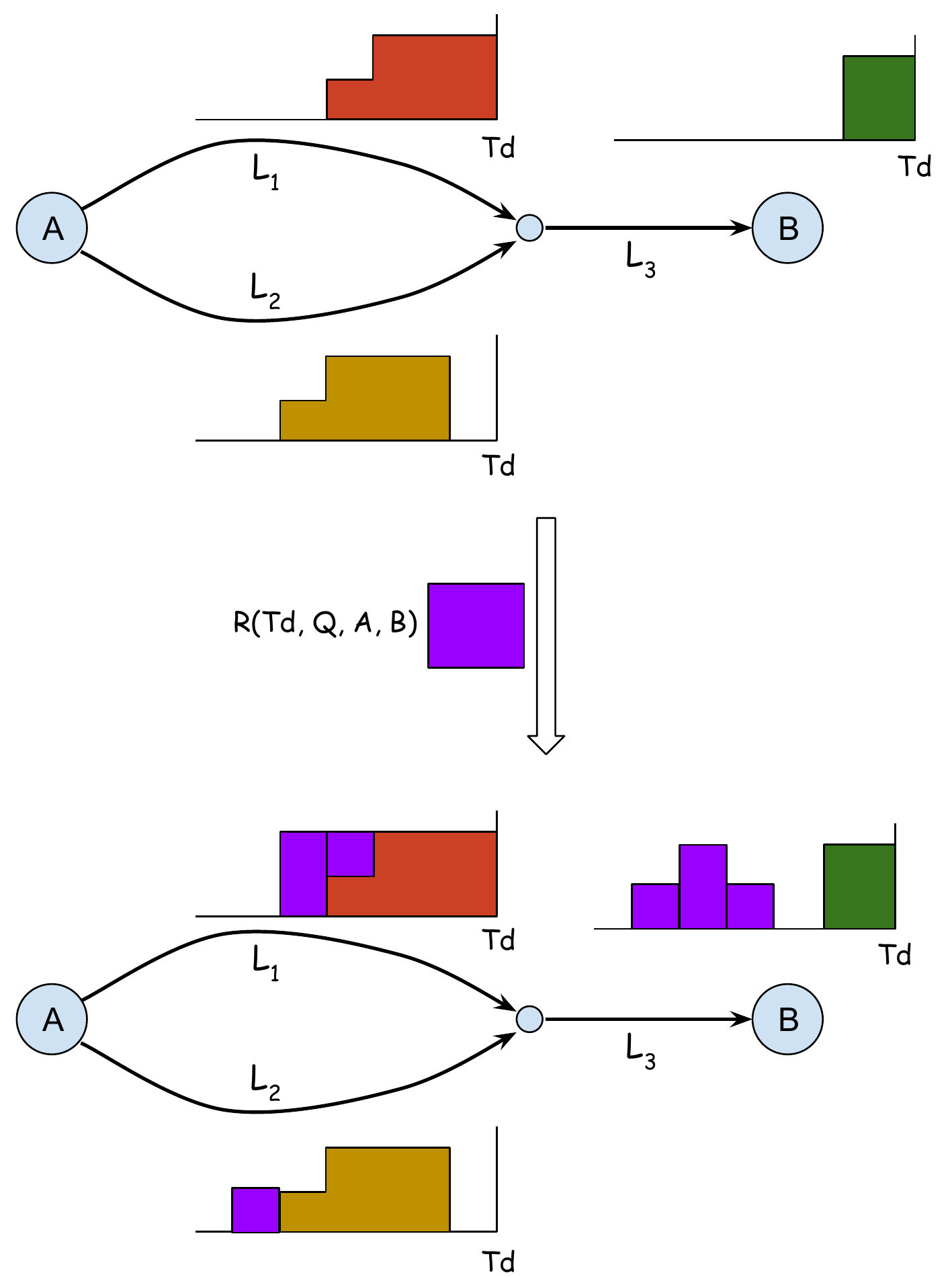}
\caption{Example of RCD allocation for network}
\label{fig:net}
\end{figure}

Fig. \ref{fig:net} shows an example of network case allocation. As can be seen, resources are not allocated as late as possible for $L_2, L_3$, however, the overall allocation cannot be pushed further into the future.

Utilizing unused resources in current timeslot by sending traffic from future timeslots is not as straight-forward anymore. A central network scheduler will have to decide how to do that. The idea is to look at the first traffic scheduled on timeline for all links, and if for some of the links, there is some unused bandwidth in current timeslot, pull some traffic from future timeslots and send it.

\begin{figure}
\centering
\includegraphics[width=0.9\columnwidth]{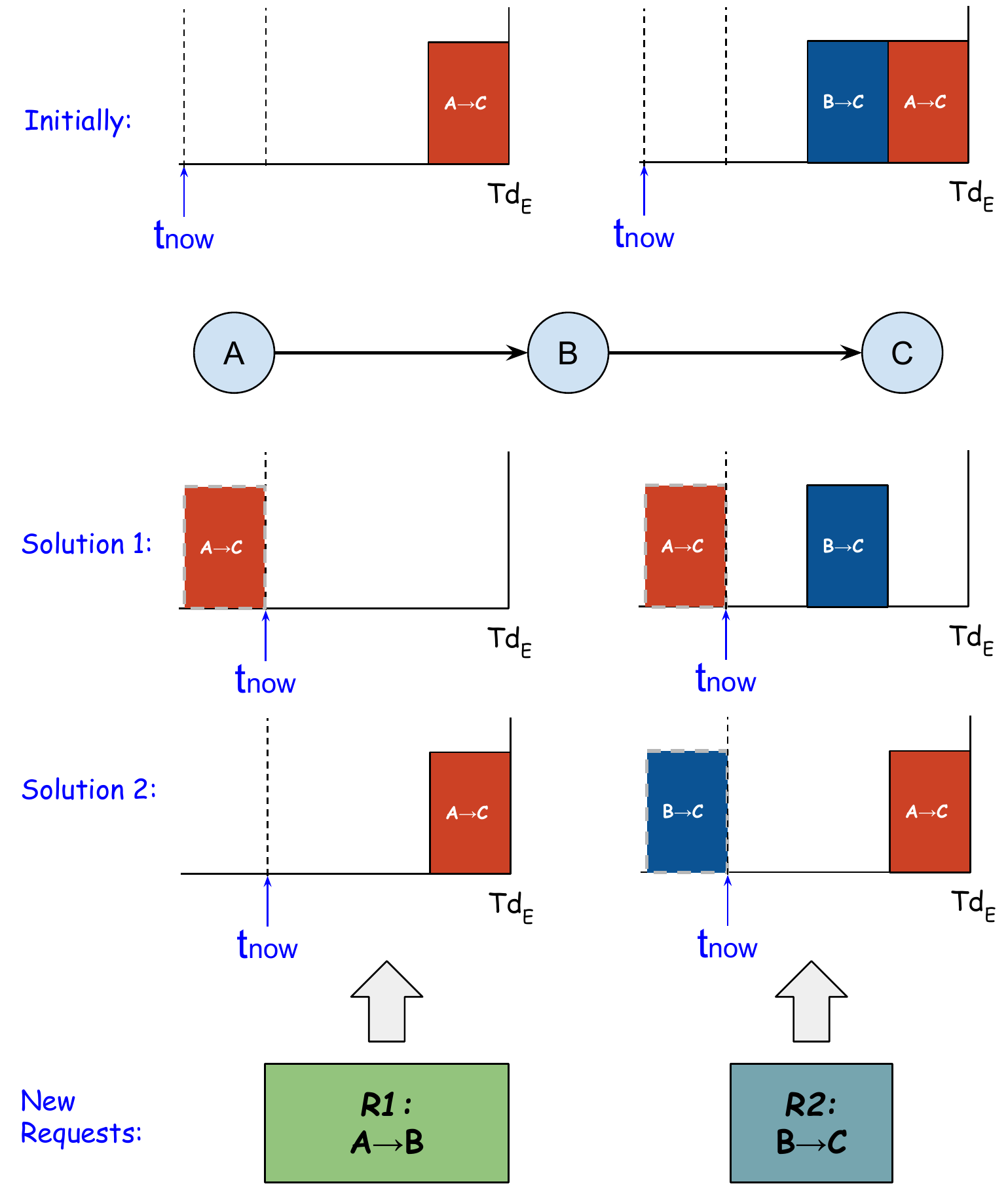}
\caption{Multiple solutions while pulling traffic}
\label{fig:net2}
\end{figure}

In this case, since there are multiple links and paths, it may not be possible to propose an optimal solution. In a single link case, all requests will have to use the same resource: bandwidth of one link. In a network case, however, requests may use resources from different links. In addition, we may not be able to pull the first scheduled request on all links and send them in current timeslot. That's because a request that is scheduled closest to current time on one link, may not be the closest to current time on next or previous link of the path. 

Fig. \ref{fig:net2} is an example of such cases. Assume $R1$ and $R2$ are two possible requests that may be submitted in $t_{now}$, however, we don't know which one it will be. At the beginning of $t_{now}$, we can either choose to go with solution 1 or 2. We prefer the first solution if $R1$ comes in, because the second solution will reject the request. For the same reason, we prefer the second solution if $R2$ gets submitted. Our lack of knowledge prohibits us from making a definite optimal decision. We plan on studying this subject further in our future studies.

\section{Conclusions and Future Work}

In this paper, we proposed RCD, a technique that makes fast scheduling of bandwidth resources possible for large transfers with specific deadlines, provides deadline guarantees, and allows for maximum bandwidth utilization. Our simulations demonstrate that for high arrival rates, RCD is up to 15 times faster than Amoeba, a system with similar objectives. RCD makes it possible to decide on request admission knowing only the residual bandwidth. We also proposed a method to reserve bandwidth for important requests with deadlines whose content is not available for delivery at current time. Finally, we proposed an idea on how RCD can be further extended and applied to a network connecting multiple datacenters. Further study of RCD considering soft deadlines, variable highpri traffic rate, application and simulation of RCD for a network case, and comparison of RCD for a network case with other schemes are identified as future work. 


\end{document}